# Nonequilibrium Kinetic Modeling of Sintering of a Layer of Dispersed Nanocrystals

**Vyacheslav Gorshkov**,[a] **Vasily Kuzmenko**,[a] and **Vladimir Privman**[b,*]

[a] National Technical University of Ukraine — KPI, 37 Peremogy Avenue, Building 7, Kiev 03056, Ukraine

[b] Center for Advanced Materials Processing, Department of Physics, Clarkson University, Potsdam, NY 13699, USA

**Abstract**

We report a kinetic Monte Carlo modeling study of nanocrystal layer sintering. Features that are of interest for the dynamics of the layer as a whole, especially the morphology of the evolving structure, are considered. It is found that the kinetics of sintering is not entirely a local process, with the layer morphology affected by the kinetics in a larger than few-particle neighborhood. Consideration of a single layer of particles makes the numerics manageable and allows visualization of the results, as well as numerical simulations of several realizations for statistical averaging of properties of interest. We identify optimal regimes for sintering, considering several particle size distributions and temperature control protocols.

**Keywords:**  sintering;  nanocrystals;  nanoparticles;  kinetic Monte Carlo

___________
[*]Corresponding author: e-mail privman@clarkson.edu; phone +1-315-268-3891

## 1. Introduction

The process of sintering has numerous applications and therefore many recent experimental efforts have been devoted to its study.[1-9] Sintering was also modelled by using several approaches[10-23] ranging from continuum theories to mesoscopic statistical-mechanical approaches. These studies have yielded information of the structure of the sintered materials, as well as on more local properties morphological properties such as neck formation and particle merging. The overall modeling of sintering would require multiscale approaches[13,16,17] connecting continuum/finite-element[14,15] and more atomistic kinetic Monte Carlo[11,16,17,23] and molecular dynamics[18] methods. However, most reported results have focused on a particular length scale of the process.

Recent experimental developments[4] have utilized capabilities of synthesizing highly crystalline nanoparticles of well-defined distributions of size, shape, and surface properties. The use of nanoparticles can improve the connectivity (conductance)[4] in sintering of layers including noble metal nanocrystals that are dispersed[2-6,8,24-28] in paste-like materials containing other additives. The crystalline nanoparticles can be uniform or distributed in their sizes and shapes, ranging from even-proportioned, approximately spherical shapes, to flake-shaped or needle-shaped particles. Better connectivity of the metal in the resulting films can potentially be obtained by a proper choice of the initial dispersion, such as the inclusion of smaller particles that snuggly fit in the voids between larger particles[4] and facilitate neck formation during the sintering process.

More generally, several properties—mechanical, density, conductance—of materials sintered by utilizing bi-modal or other size distributions at various mixing conditions, have been studied for some time.[4,12,19,21,23,24,29,30] Recently, we initiated the development and application of a kinetic Monte Carlo (MC) methodology[23] which, while requiring a substantial computational effort, enables a semi-quantitative study of neck formation potentially leading to merger (sintering) between the same and different-size nanoparticles and groups of nanoparticles in settings mimicking nanoparticle sintering. Assuming SC lattice structure,[23] mechanisms of neck initiation via clustering or layering leading to bridging between nearby nanocrystaline faces of pairs of closely spaced nanoparticles in several few-particle configurations were explored. Conditions, such as the local geometry and temperature control for stable bridging and merger of nearby particles were identified.

In this work we study aspects of sintering which are not just local and involve processes in layers of nanocrystals rather than in small clusters. We consider FCC crystal symmetry which is appropriate for noble metals. Large-scale numerical simulations are advanced to allow consideration of sintering of approximately 200 nanocrystals of possibly varying sizes, with their centers in a single plane. The aspects of sintering that can thus be studied include the following. We find that the kinetics of bridging is a collective process, with the local neck formation affected by the kinetics in a larger than few-particle neighborhood. Consideration of a single layer of particles makes the numerics manageable and allows for an easy visualization of the results, as well as simulation of enough realizations for statistical averaging of certain properties of interest. We also attempt to identify optimal regimes for sintering as far as particle size distribution and temperature control are involved.

Sintering involves nonequilibrium dynamics[21,23,25-35] of kinetic processes of transport of matter: atoms, ions or molecules, to be termed "atoms" for brevity. These processes include on-surface restructuring, and detachment/reattachment. Here such processes are modelled within the kinetic MC approach suitable for describing the dynamics of the surface and shape morphology in sintering starting with nanocrystals of linear sizes of order 20 atomic layers. The model was recently developed[36] for growth of nanoparticles of well-defined shapes. It was also applied to formation of on-surface structures of interest for catalysis.[37,38]

A review of modeling approaches to sintering[35] emphasizes the fact that presently theoretical approaches are qualitative, at best semi-qualitative in their predictive capabilities. Indeed, as mentioned sintering requires a mutiscale approach[11,16,17] that is presently not available. What is needed is parameter extraction from microscopic modeling for mesoscopic description, and then connecting the mesoscopic models to continuum modeling at larger scales. Presently, typical models of sintering[10-23,32-35,38-41] consider only specific scales of the process. Our kinetic MC model was originally developed[23,36] for the regime in which well-defined nanoparticle shapes are known to emerge[36,42] in situations when particles are grown by plentiful external supply of diffusing solute matter atoms. The model is therefore mesoscopic, with the microscopic parameters introduced via temperature-dependent Boltzmann weights, as described in the next section, which describes the theoretical model. For sintering, the matter is not supplied externally but is rather transported between and within the closely packed, ultimately fully or partly merging particles.

In the present work, we consider a more realistic (for the experimentally relevant case of noble metals) FCC lattice structure and invest a significantly larger computation effort than in the earlier study that introduced this kinetic model for sintering,[36] surveyed in Sec. 2. This allows us to obtain new results that address the following issues of interest in sintering. We explore the role of temperature control and of nonuniformity of particle size distribution in sintering of two-dimensional layers. Section 3 reports our results and offers a discussion, including the conclusions that the properties of sintering are not just local but have certain collective aspects, as mentioned earlier.

Modeling of layers also sets the stage for future work on three-dimensional geometries, which would require even larger computational resources. Furthermore, studies of layer sintering as a collective phenomenon are of interest on their own, because recent experimental work offers an example[9] of a complex-structured film of interest in device structure design, that sinters with the final morphology and material composition varying in thin layers. This "speciation within the device layers" was not fully explained,[9] but can likely be attributed to the interplay of the variation of the temperature control and matter transport within the film thickness.

## 2. Description of the Theoretical Approach

The present kinetic MC methodology was tested in modeling nanocrystal synthesis, surface synthesis of nanostructures, and aspects of sintering.[23,36-38,42] Here we outline the model with emphasis on those implementation details which are specific to the present work. We assume that atoms (or ions, or molecules) are initially all in nanocrystals. As the sintering process goes on, they can detach, diffuse as single entities, reattach to the original or other particles, as well as hop on the particle surfaces. The initial particles have the FCC lattice symmetry without defects, and are all registered with the underlying FCC space-covering lattice, of cubic-cell lattice spacing $a$. The detached single atoms undergo off-lattice diffusion modeled as hopping at random angles, for each atom per each MC time step (to be specified later), in steps taken for convenience as $a/\sqrt{2}$ (equal the primitive-cell lattice spacing of FCC).

Diffusing atoms can be recaptured at vacant lattice positions nearest-neighbor to the particles. Each vacant site is surrounded by its Wigner-Seitz cell. An atom that hops into such a cell is captured and positioned registered exactly at the cell's center. The on-surface restructuring, described below, also preserves the "registration" property. Attempts to hop into cells that are filled are rejected. Atoms that are in particles can move on the surface or detach. The registration of the attached atoms is important[23,36-38] for maintaining nanocrystal morphologies of relevance. This imposed rule precludes the formation of structure-spanning defects that can dominate the dynamics of the shape/feature variation of nanocrystals as a whole, by preferentially driving the growth of selected crystalline faces and/or by sustaining unequal-proportion shapes. Nanocrystal shapes considered here, motivated by the experimental work,[4] are synthesized in the regime of approximately equal-proportion (so called "isomeric," nearly spherical) growth that can be disrupted by large defects. Such defects are dynamically avoided/not nucleated at the microscopic scale, and this property is *emulated* by the exact registration at the mesoscopic scales of our modeling.[23,36-38]

Initially, particles are positioned with their centers in the plane, randomly distributed in an approximately square domain containing about 200 particles for typical realizations studied (see Sec. 3). An illustration of a typical initial configuration is given in Fig. 1, which also shows later-time sintered configurations for two different temperature-control protocols (specified in Sec. 3.2). The particle assembly is then enclosed in a curved-surface box with reflecting boundary conditions for diffusing atoms. This box is defined by points at a distance of $6a$ vertically from the extremal atom of each particle above, below, and also horizontally on the outer sides of the assembly. The box was then smoothed out by using a two-dimensional spline, in order not to create artificial voids near smaller particles in the distribution. Additional explanations are given in the next section, where we offer illustrations and describe how the initial particle shapes and size distributions were selected.

Let us now outline the rules of the kinetics of atoms that are in particles (i.e., are parts of connected groups of two or more crystal-lattice-registered atoms). These atoms can detach and also hop to their nearest-neighbor vacant lattice sites, the latter without detaching from the cluster. The probabilities for each atom that is not fully blocked by neighbors to move are proportional to Boltzmann factors that introduce the temperature dependence. During each unit time step, here a MC sweep through the system, in addition to randomly moving each diffusing (detached) atom once of average, we also attempt to randomly move each attached atom that has vacant neighbor site(s), once of average. For an atom in the latter category, with a coordination number $m_0 = 1, ..., 11$ (for FCC), we take the probability for it to actually move during a time step as $p^{m_0}$. This can be interpreted as representing a free-energy barrier, $m_0 \Delta > 0$, were $p \propto e^{-\Delta/kT} < 1$. If the atom actually hops, it will then move to one of its $12 - m_0$ vacant neighbor sites with the probability proportional to $e^{m_f|\varepsilon|/kT}$ (normalized over all the targets). Here $\varepsilon < 0$ is the free-energy measuring binding at the available target site(s), the final coordination of which will be $m_f = 1, ..., 11$ for hopping, and $m_f = 0$ for detachment.

To recapitulate, hopping of connected atoms occurs with surface diffusion coefficient related to $p$, involving a free-energy scale $\Delta$, such that

$$p \propto e^{-\Delta/kT}. \qquad (1)$$

In addition, another free-energy scale, $\varepsilon$, is defined by the local binding, and here it will be varied via the dimensionless parameter

$$\alpha = |\varepsilon|/kT. \qquad (2)$$

Earlier studies[23,36-38] suggest that the appropriate regimes of typical FCC nanocrystal morphologies in nonequilibrium processes are qualitatively represented as follows. We can take the reference values $\alpha_0 = 1$ and $p_0 = 0.7$, and then increase or decrease the temperature by decreasing or increasing $\alpha$, respectively, here varied in the range between $\alpha = 0.7$ to $2.0$, with $p$ appropriately adjusted as

– 6 –

$$p = (p_0)^{\alpha/\alpha_0}. \tag{3}$$

This *assumes* that both energy scales ($\Delta$ and $|\varepsilon|$) remain approximately constant in the considered temperature range, and we also comment that we do not account for a possible temperature dependence of the diffusion constant of detached atoms.

## 3. Results and Discussion

### 3.1. The Initial Configuration

Let us first describe aspects of the present sintering model that are important for our selection of the initial setup of the system, i.e., the initial configurations at $t = 0$. We note that nanoparticles of interest here are typically synthesized[4] under the nonequilibrium conditions of plentiful supply of atom-size matter. Earlier studies indicate[23,42] that their shapes are then bound by lattice planes of symmetries similar to those in the equilibrium Wulff constructions,[43-45] but with different proportions. However, sintering is usually carried out under the conditions whereby the gas of diffusing atoms is not externally supplied but is rather originating by detachment from the original particles. Earlier work for SC symmetry resulted in the following observation,[29] confirmed by our modeling for few-particle groups with FCC symmetry, which is not detailed here: Particles not only develop connecting necks and merge, but the resulting flow of matter by various mechanisms discussed in the preceding section also causes their outer shapes to round up (while retaining their crystalline cores).

Another property observed for few-particle SC configurations[29] and confirmed for FCC has been that neck formation between truly nanosize crystal surfaces that face each other, can proceed by either layering or clustering mechanisms, and, most importantly, the time scales of this initiation of bridging significantly vary/fluctuate depending on the local geometry and between different MC realizations. One way to speed up the process is to have small inter-particle gaps. Experimentally, recent work on sintering[4] that observed improved film properties with distributed particle sizes has reached similar conclusions. The particle sizes were distributed assuming approximately near-spherical particle shapes, aiming at as snug a fitting of the smaller particles in the voids between the larger ones as possible, and generally emphasis was put on as close a proximity of the particles to each other as possible.

However, sintering of noble-metal nanocrystals is a complicated process. Particle synthesis leaves organic residues at their surfaces, such as Arabic gum. The particles are then mixed into a viscous paste that contains other fillers and is printed as ink on a substrate. The full preparation process involves mechanical shaking/tapping/compression at various stages before the sintering. The latter typically is combined with firing which burns away some of the organics.[25-28] Experimental data on a pre-sintering tapped/compressed density for nanoparticles[25] suggest that an assumption of inter-particle gaps of order 3 to 5 atomic layers is more realistic than direct contact, largely due to the presence of the stabilizing organics.

Given all the above observations, and considering the need to make our simulations as fast as possible in terms of the initiation of sintering in order to allow us to study large enough layers (as mentioned, up to 200 particles), we devised the following procedure for the initial configuration selection. Our plane of nanoparticle centers was for simplicity selected as (100) type, and the particles of various sizes were thus registered in their orientations. The particles were cut out of the FCC lattice as approximately spherical, of diameter $D$, by keeping only the unit cells with their centers within such a spherical volume. The size distribution was taken Gaussian, proportional to

$$\exp\left[-\frac{(D-D_0)^2}{2\sigma^2}\right], \quad \text{with } D > 0, \tag{4}$$

were, as mentioned earlier, $D_0 = 10a$, and we used $\sigma = 2a$. This corresponds to a single-peaked distribution which is not overly sharp,

$$\sqrt{\langle \Delta D^2 \rangle} \approx 0.2 D_0. \tag{5}$$

This choice is further discussed later.

A separate procedure was devised to generate the initial planar particle configuration which is dense enough, by bringing the particles to the outer boundary of the previously positioned particle cluster one at a time, with the initial near-contact (minimal-gap approach). The first couple of particles were placed in near-contact randomly, with minimal gaps between them. The $n$th added particle (for $n > 3$), was then slid around the full periphery of the assembly to the location, $\vec{r}_n$, that minimized the quantity

$$\sum_{i=1}^{n}(\vec{r}_i - \vec{r}_{\text{av}})^2, \quad \text{where} \quad \vec{r}_{\text{av}} = \sum_{i=1}^{n}\vec{r}_i/n. \tag{6}$$

This nonlocal compacting avoided formation of voids, such as, e.g., in the diffusion-limited growth that leads to fractals.[46-48] Our procedure yields coverage with small gaps only, with the pre-set minimum distance between the closest particles. This was then used as the initial configuration. For example, the configuration shown in Fig. 1(a) has about 200 particles with the minimal gaps of $2a$, up to small variations imposed by the cellular structure of the particles truncated within spherical regions. We designate such initial configurations as Type I. In addition to averaging over Type I configurations, generation of each of which took a significant computational effort, we also then used an ensemble of denser configurations, designated Type II, obtained by fattening each particle by adding atoms in radial layer of size $a/2$, in order to shrink all the minimal gaps down to the sizes of approximately $a$. This is illustrated in Fig. 2(b).

### 3.2. Selection of the Temperature Control Protocol for Sintering

As mentioned in the introduction, we aim at considering the effect of temperature-control protocol selection of the sintering process, and also the implications of using distributed particle sizes. This is considered in the present and the next subsection. For the latter, we also generated initial configurations with uniform particles, of diameters $D = 10a$. We note that in applications the aspect of the sintering process that is the easiest to control involves the temperature variation. Usually, the temperature is elevated for some time to cause sintering. There are important tradeoffs involved in selecting the appropriate protocol. Faster variation makes manufacturing more efficient. However, it requires reaching larger temperatures that can damage non-metal components in the film.

Figure 3 offers examples of the temperature control protocols used in our numerical experiments. These are designated as A07 and B07 for the short-duration and long-duration higher maximum temperature case, and A09 and B09 for the lower maximum temperature choices, see Fig. 3 for details. Usually, the heating can be quite fast, and therefore we generally took somewhat faster heating pulses than the cooling processes. Most of the actual changes in the system occur at the larger temperatures.

We note that our initial configurations are a type of the two-dimensional random close packing, with coordination number of 4. We know from previous work[23] that locally, smaller particles can facilitate bridging, but they can also be ultimately dissolved into larger particles thus resulting in voids. High-quality sintering will therefore be a process during which those particles that did not yet lose their identity keep as close to 4 contacts as possible with the surrounding more globally sintered structure. Note that the upper number of contacts for three dimensions will be 6 (typical for three-dimensional random close packing).

Figure 4 illustrates the fact that a short pulse, to a higher temperatures, provides an optimal protocol for temperature control in order to allow a reasonable number of particles to establish more contacts and retain them as long as their identity is not entirely erased. Figure 5 shows an example of how the same initial particles develop less contacts for other temperature control choices, including lower temperatures and/or slower pulses. These images are just illustrative: We found such trends for a larger number of realizations studied.

This observation has an interesting practical implication. It transpires that for good-quality sintering one cannot avoid elevating the temperature. However, making the high-temperature pulse shorter can actually improve the quality of the bridging. This effect is nonlocal, because elevated temperature can cause not only bridging but also flow of matter especially from smaller particles to larger ones, the latter as long as the particles are very close to each other or connected. Some of the smaller particles' matter will then recede in the directions of their formed connections or near-contacts, and for prolonged heating they will not form new necks to initiate additional bridging. Effectively, longer heating makes more particles "small" as far as retaining their identity for long times goes.

Smaller particles can dissolve after establishing few bridges (or by evaporation) even for the fast heating case. This is illustrated in Fig. 6, in which we identified a fragment of an initial configuration that has a smaller particle surrounded by larger ones. Not only was it dissolved, but it actually left behind a void in the resulting structure, Fig. 6(e).

### 3.3. Optimization of Temperature Control

In the preceding subsection we considered specific snapshots of configurations during the sintering process as numerically simulated in our model, illustrated in Figs. 4, 5, and 6. In order to consider the statistics of optimizing the temperature control for the system as a whole, we need to first discuss the limits of applicability of the present model for larger time scales of the process. Later stages of sintering are accompanied, in our model, by the development of voids in the structure, as can be seen in the snapshots in Fig. 4, 5, 6, and more globally in Fig. 1(c). However, in real systems the sintered structure compactifies and new effects are expected that are not included in our mesoscopic modeling approach.

Indeed, we already emphasized that sintering is a multiscale process. It is well known experimentally that at later times the sintered systems will shrink, typically losing up to order 25% of their volume. A continuum macroscopic description of the mechanical transport of material will be needed to properly describe this process. Our model simply does not incorporate such material flow, though earlier work has found[23] that the developing large voids in our MC simulations can indeed account for up to 30% of the total system volume being in the voids at large times. Furthermore, it was also noted[23] that at later stages of the sintering process the gas of detached atoms only contains approximately 1% of the total matter in the system, confirming that most of the dynamics occurs by the flow of matter in the connected structure, again emphasizing the need for continuum modeling for large times. Furthermore, continuum description can address additional surface restructuring[49] and larger-scale mechanical effects such as structural movement and effective forces between connected structure parts as driving the dynamics.[10,50]

To recapitulate, structural changes expected for large times cannot be considered within the present model, and because these changes can modify the gaps in the structure, they can change the dynamics of sintering. Therefore our study reported below of the temperature control optimization applies only as long as the voids are not too large or the materials of the voids (such as the added organics) are not depleted (for instance by the accompanied firing mentioned earlier). Typical later-stage configurations for the Type I initial state selection are illustrated in Fig. 7. The time scales for our temperature control protocols, Fig. 3, were selected in such a way that at the end of the process the voids are still relatively small as compared to the overall sample dimensions (Fig. 7) and therefore it is hoped that our conclusions described below apply semi-qualitatively for the considered stage of the sintering process.

The quality of sintering can be defined by different criteria, such as the mechanical or electrical (conductance) properties of the final product. However, for our mesoscopic model it is appropriate to limit the consideration to geometrical features. Specifically, long-duration temperature pulses lead to larger voids, as seen in Fig. 7(c-d). We already commented that because of the orders-of-magnitude difference in times required for bridging between crystal faces of various orientations and separations, slow heating pulse allows densification of local sintered structures at the expense of the rest of the material. The latter is then more prone to being sucked into the more sintered regions, leaving behind larger voids. For faster temperature control protocols, we note that the lower-temperature case A09, Fig. 7(b), had more than a single large connected structure. However, this varies between MC runs, and such a large-cluster count is anyway not an obvious quality measure for sintering.

More appropriate indicators of the quality of sintering would be measures of the size of the voids and connected lattice regions (not considering the detached, off-lattice atoms). For protocols A07 and A09, Fig. 8 summarizes statistics of the simplest measure of this kind: the counts of the various linear void sizes and occupied-span sizes along the *x* and *y* in-plane lattice rows, averaged over several MC realizations. Figure 8 indices that indeed, in this case most voids remain rather small, whereas the distribution of the connected-span sizes develops a noticeable large-value (as compared to the original mean particle diameter, $D_0$) tail, as highlighted in the figure. Not surprisingly, there are noticeably more such large connected spans, at the expense of the small-span part of the distribution, for the larger-temperature protocol A07. This confirms our earlier observation that the use of a fast, short-duration temperature pulse is optimal for sintering at least at the scale of mesoscopic morphology, because one then obtains better sintered (larger in linear dimension) structures without sacrificing uniformity.

### 3.4. Effects of the Initial Configuration

The initial configuration of particles can depend on the desired and available amount of the sintered material (the latter for noble metals in many cases constrained by cost), as well as on the preparation methodology. The preparation in turn involves the particle size distribution and also the pre-sintering mixing and mechanical processing that affects the relative positioning of the particles. We will address several aspects of the initial configuration choice by studying its effect on the sintering process.

Let us first consider the role of the density of the particle packing (which requires more material and can be more costly). We repeated the study of the effect of the temperature protocols on sintering for the denser Type II initial configurations prepared as described earlier. Figure 9 illustrates sintered configurations for time-dependent temperature control protocols A08 and B08 in our notation. Short-duration pulse (A08) and long-duration pulse (B08) times were the same as before, see Fig. 3. We study this temperature protocol with $\alpha_{\min} = 0.8$, in addition to the values 0.7 and 0.9 considered earlier, because we found indications that for denser initial configurations lower peak-temperatures suffice, without sacrificing good quality of sintering. This is illustrated by the bars in Fig. 10(a) for the fast heating/cooling case: The void sizes are studied for protocols A07, A08, A09. There is practically no difference in the distributions for larger void sizes for A08 vs. A07 (and actually also A09).

The same figure, Fig. 10(a), also shows the void size distribution for the high-temperature slow-pulse protocol B07. We note that the large void counts in this case notably exceed large void numbers found for all the short-pulse protocols. Furthermore, the figure also illustrates that a good fraction of this excess of large voids was formed already during the heating part of the slower protocol. The rest developed during cooling. On the other hand, for the fast protocol A07, Fig. 10(b) shows that there is no additional creation of large voids, while a notable fraction of small voids are eliminated during the cooling stage.

The overall observation suggested by the presented results and our other studies (with various parameters, protocols, etc.) is that the advantages of the use of short-pulse protocols of temperature control, reaching sufficiently high temperatures, are confirmed. To the extent possible, denser initial configurations should be preferred for quality sintering with lower peak temperature required.

Let us consider the effects of sharpness of the initial particle size distribution on the quality of sintering. We note that our choice of the Gaussian distribution described in Eq. (4), with not too large a dispersion, see Eq. (5), ensured that the generated dense configurations, such as Fig. 1(a) and 2, had the morphology of two-dimensional random close packings. Experimental work included consideration of bimodal distributions[4] for which it was found that small particles fitting between approximately uniform spherical larger particles (for this, the radii were taken in proportion of approximately 1:7) can improve the conductance of the sintered noble metal films. The present single-particle-layer modeling sets the stage for considering such systems, but three-dimensional simulations will be required for a proper placement of large and small spheres to obtain the expected configurations of relevance. We plan such studies in the future, within the context and with the optimal protocols and parameter values identified in the present study for two-dimensional geometry.

One aspect of the problem that can be addressed with the two-dimensional modeling is the effectiveness of trying to sharpen the particle size distribution to achieve better initial particle coordination. Let consider the case of uniform particles (diameters $10a$) for which our procedure for generating the initial configurations will result in a hexagonal packing with coordination number 6. This is illustrated in Fig. 11(a). Based on our earlier considerations, let us check the extent to which the sintering process for the earlier identified close-to-optimal conditions, specifically: dense initial packing with minimal particle distance of approximately $a$, and with the fast-pulse protocol A08, is affected by the use of the uniform particles. The evolution of a possible initial configuration is illustrated in Fig. 11. Comparing the final state to that of Fig. 9(b), we note the presence of rather narrow but long voids, which can actually degrade certain mechanical and possibly conductance properties when uniform particles are used. Even though our measure of the void sizes focuses on the linear chains of unoccupied lattice sites, we note that it offers a quantification of this visual conclusion by yielding a larger count of several-lattice-site empty spans, Fig. 12. Note that the data in Fig. 12 represents averaging over 20 different MC runs that also differed by their initial configurations that can have several possible orientations of the hexagonal pattern. These orientations are set by the first couple of particles' (which are truncated to spheres of cell centers) placement registered with the lattice, which also results in small fluctuations of the minimal interparticle distance.

Figure 13 offers snapshots of peak-temperature and ending-time configurations for slow protocols at two different temperatures, both started from the same random uniform-particle configuration. We again see the earlier noticed property that gap sizes are additionally increased for long-pulse protocols during the cooling stage.

### 3.5. Conclusion

When applied to two-dimensional layers, the kinetic MC model of sintering leads to several interesting observations. We obsevred that a short-pulse heating protocol to high enough peak temperature is the optimal for avoiding large gaps in the resulting structure. The initial particle size distributions considered, which were chosen not overly poly-disperse, allowed for dense initial packing. However, there was no obvious virtue in seeking truly mono-disperse (uniform) particle size distribution, because quality of sintering is not further improved. We also found that sintered configurations have certain features of their morphology that are obviously not entirely local but span mesoscopic sizes notably larger than few-atom, and which depend on the dynamics of the system as a whole, as well as on the initial configuration.

Therefore, mesoscopic modeling of the type reported here can offer useful qualitative and potentially semi-quantitative information of the sintering process. However, ultimately the task of combining various-scale approaches within a multi-scale description will have to be addressed, which, to our knowledge, remains an open challenge in the theory of sintering generally. This includes both the "parameter extraction" from truly microscopic models for our and similar mesoscopic approaches, as well as connecting the latter to the more continuum larger-scale dynamics,[10,49,50] such a particle motion, flow of matter, and also surface restructuring during sintering.

The present simulation was large-scale, requiring consideration of systems of up to $4\times 10^5$ atoms. Simulations were carried out using parallel-processing clusters with CPUs such as Intel® Core™ i7-870, 2.93 GHz, or Intel® Xeon® X5660, 2.80 GHz. They took 1-2 days of CPU per each run, depending on the temperature control protocol, the initial configuration, and other system parameters. For gathering all the reported statistics (see Figs. 8, 10, 12), the total amount of CPU was equivalent to approximately 4 months on 60-70 cores running in parallel. The MC approach was validated in earlier reported studies[36] of the selection of nanocrystal shapes in synthesis. It could reproduce all the experimentally observed[51] metal nanocrystal shapes reported (those experiments and modeling were for BCC). Future studies within the present mesoscopic kinetic MC model will include three-dimensional particle arrangements, limited perhaps only to few particle layers due to computational resources required, to allow for proper geometry for bimodal distributions that involve small particles fitting in the voids of stacked layers of larger particles.

## Acknowledgements

The authors thank Prof. D. V. Goia and Dr. I. Sevonkaev for rewarding scientific interactions and collaboration, and acknowledge the use of the computational cluster infrastructure of Karpenko Physical-Mechanical Institute of the National Academy of Sciences of Ukraine (NASU).

**FIGURES**

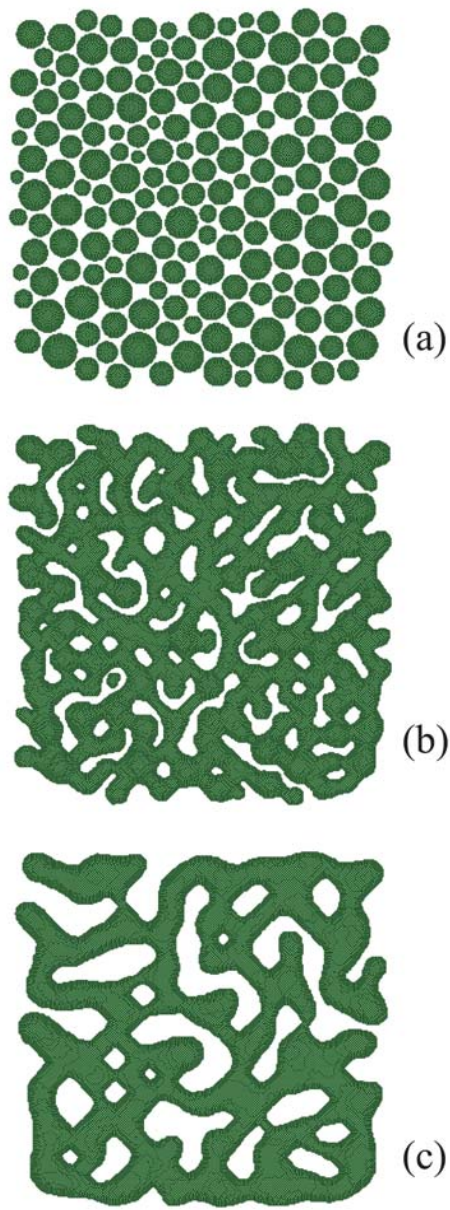

**Figure 1.** (a) Top view of a typical initial particle configuration, offered here for illustration purposes only. (b) The connected clusters after sintering involving a linearly-increasing followed by a linearly-decreasing temperature-control protocol A07 (specified in the text, see Sec. 3.2). (c) Same as (b) but with longer-duration temperature increasing and decreasing stages of the process, protocol B07 (see Sec. 3.2). Note that here and *in all the other images* the detached atoms are not shown.

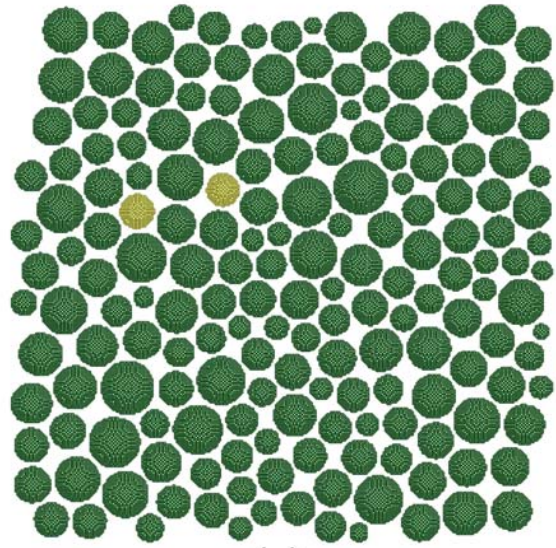

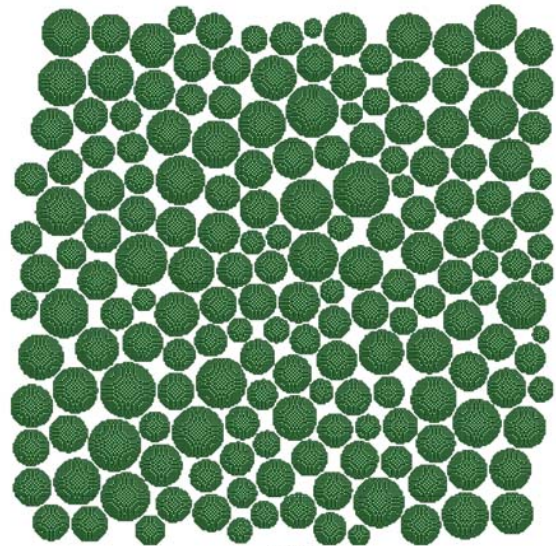

**Figure 2.** (a) This initial configuration, of Type I, was generated by the procedure outlined in Sec. 3. The region shown is of the size $150a \times 150a$. The two particles drawn in yellow are those that will be individually followed during sintering. Note that Fig. 1(a) corresponds to a different initial realization with the same parameters for the particle size distribution. (b) This Type II configuration was obtained from that shown in panel (a) by fattening each particle as described in the text. As a result, the minimum gaps between the particles were reduced from approximately $2a$ to $a$ (see text), and the total number of atoms in the resulting Type II configuration is about 15% larger than in the original Type I configuration.

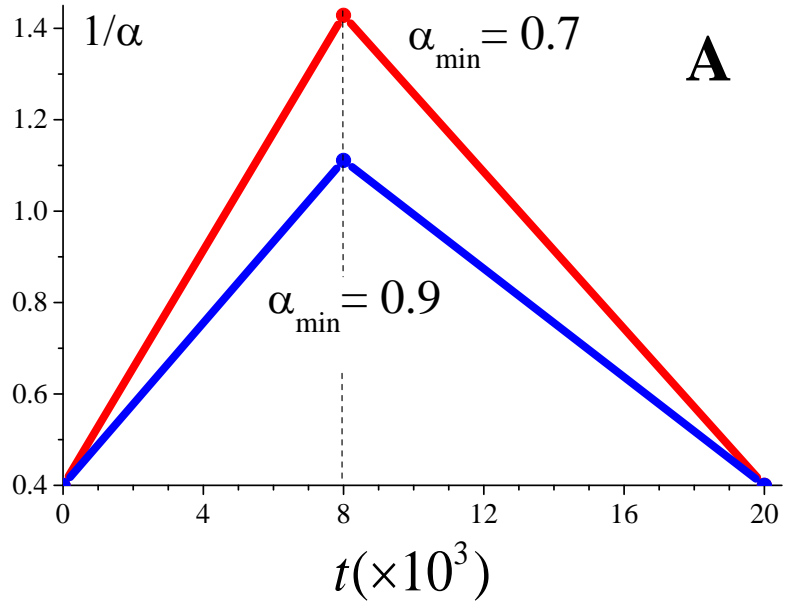

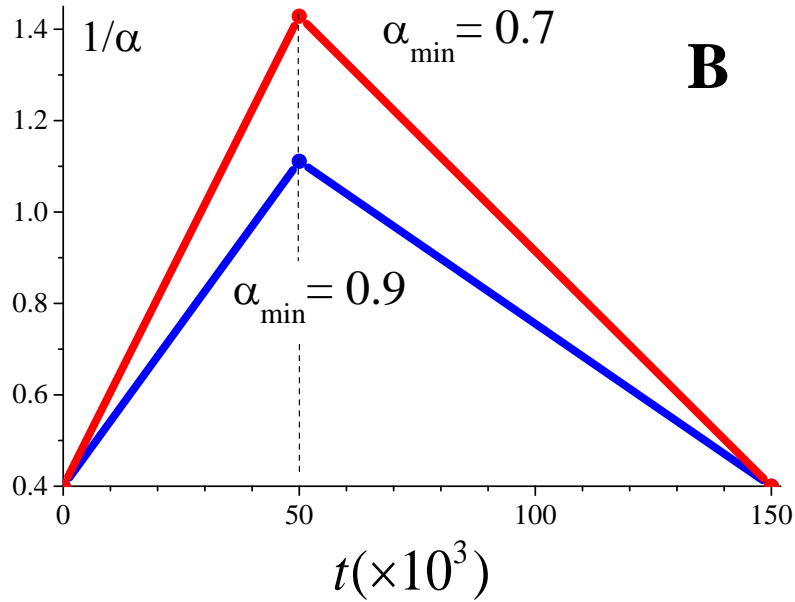

**Figure 3.** Illustration of protocols used in our numerical experiments to linearly vary the temperature up and then down. **A**. Typical short-duration protocols. The temperature is proportional to $1/\alpha$, and it reaches its largest values at the shown $\alpha_{min}$. The lowest temperature corresponds to $\alpha = 2.5$. We denote these protocols A07 (for $\alpha_{min} = 0.7$) and A09, in a self-explanatory notation. **B**. Typical long-duration protocols to be denoted B07 and B09, in a similar self-explanatory notation.

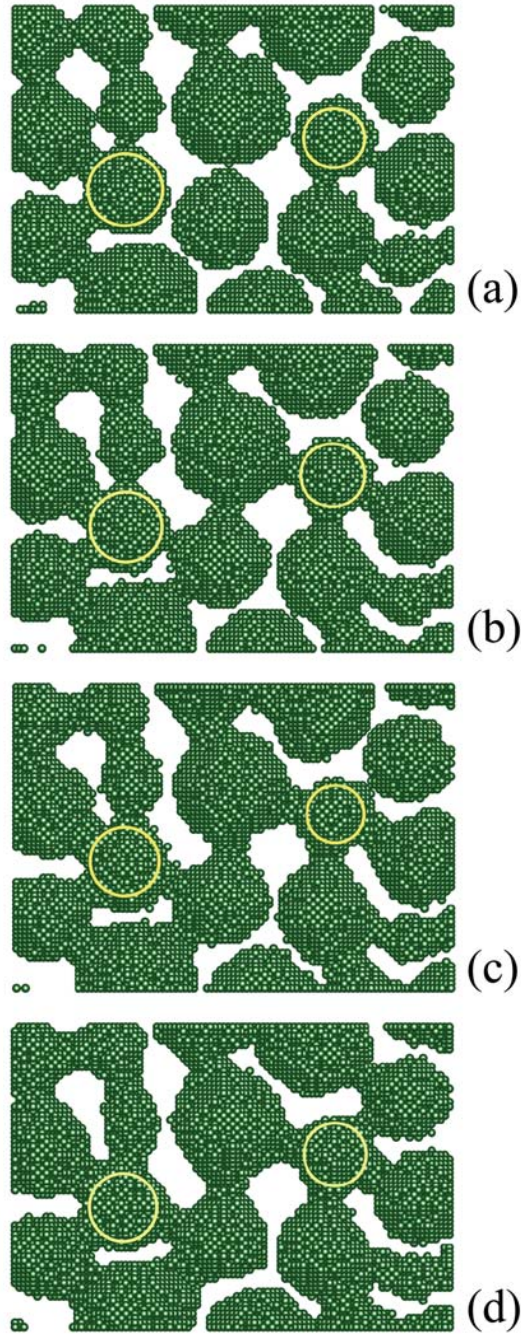

**Figure 4.** Fragment of the configuration during sintering starting with the initial state shown in Fig. 2(a), for $t = 6.0, 6.5, 7.0, 7.5 \times 10^3$ MC time steps, shown in panels (a), (b), (c), (d), respectively, for the temperature control protocol A07 (short pulse, to higher temperature). The particles highlighted in yellow in the initial configuration, see Fig. 2(a), are now connected but retain their identity and are marked by yellow circles. They ultimately establish four contacts and retain them for larger times.

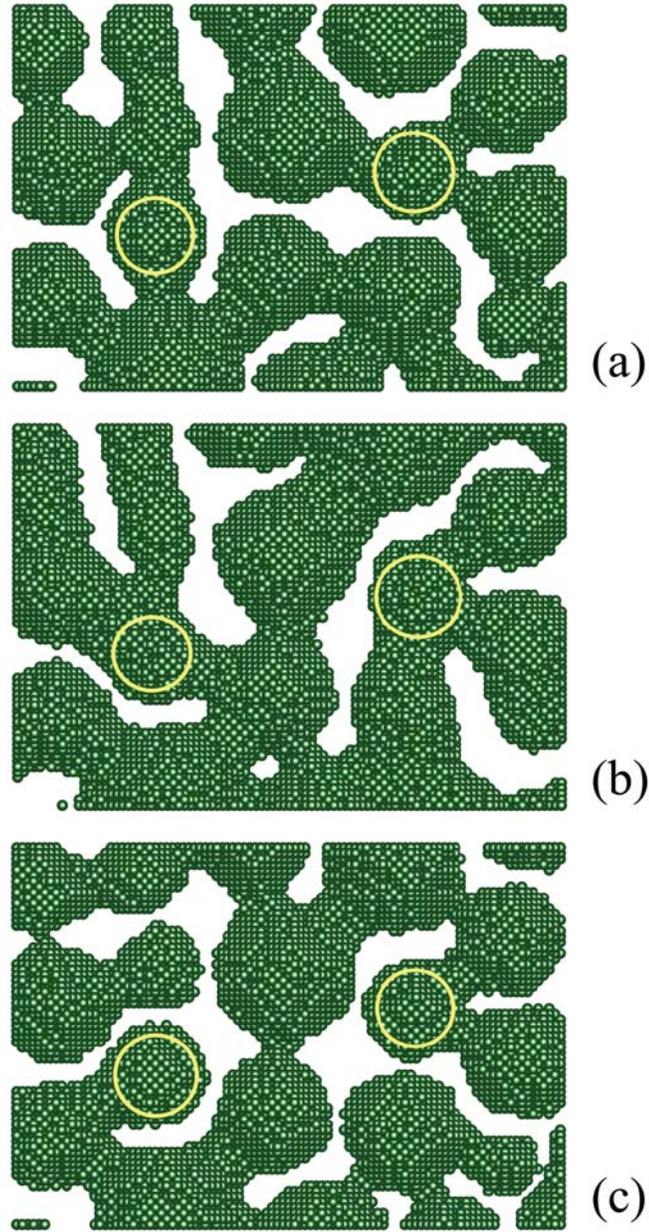

**Figure 5.** The same fragment as in Fig. 4, of the configuration during sintering starting with the initial state shown in Fig. 2(a). (a) For $t = 12 \times 10^3$ MC time steps, for the temperature control protocol A09 (short pulse, to lower temperature). (b) The same for $t = 50 \times 10^3$ MC time steps, for B09 (long pulse, to lower temperature). (c) Here $t = 30 \times 10^3$ MC time steps, for B07 (long pulse, to higher temperature). The illustrated times are those for which the particles highlighted in yellow in initial configuration developed contacts as seen in the images. These particle are marked by yellow circles. For larger times, they retain the same contacts, but do not establish new ones.

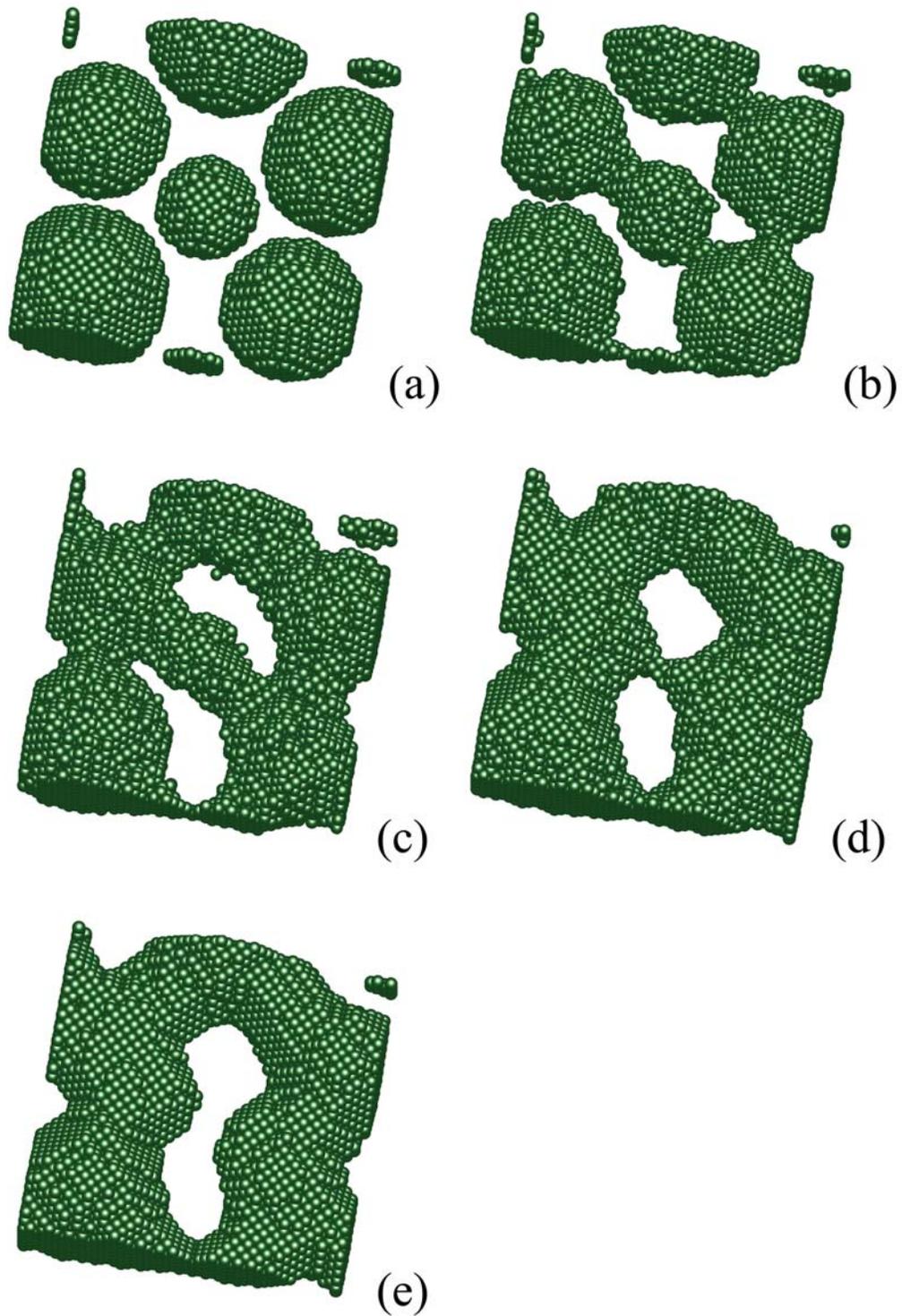

**Figure 6.** Fragment of a configuration during sintering starting with the initial state shown in panel (a). Sintering with the temperature protocol A07 leads to the consecutive configurations (b), (c), (d), (e), for $t = 5.7, 8.0, 14, 15 \times 10^3$ MC time steps, respectively.

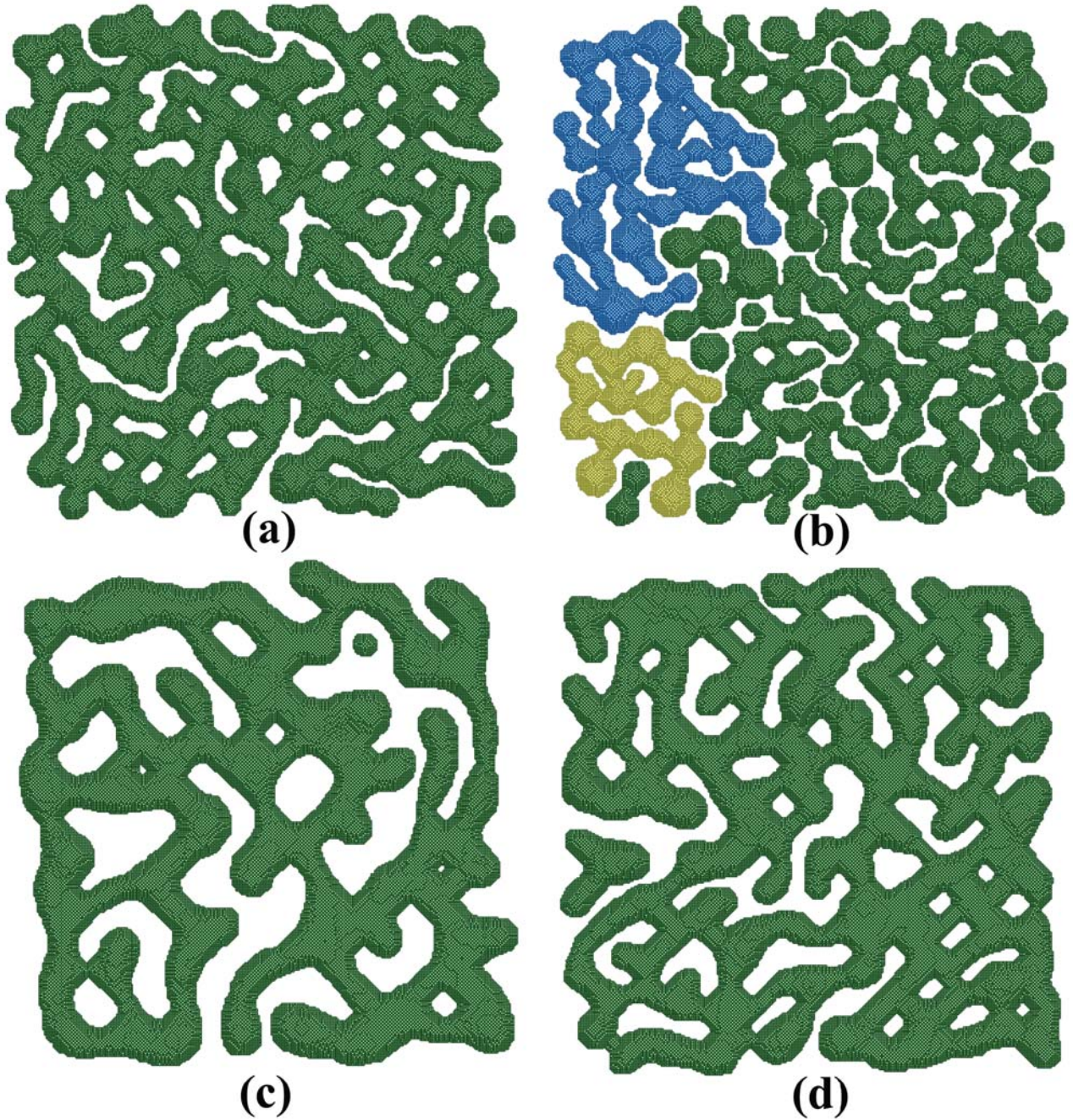

**Figure 7.** Typical later-stage sintered structures obtained at the end times (defined in Fig. 3) for different temperature protocols, starting with the Type I initial particle configuration shown in Fig. 2(a): (a) A07; (b) A09; (c) B07; (d) B09. Note that for protocols A07, B07, B09, the connected structure contains most of the matter except for some small separate structures (the gas of detached atoms, which are not shown, is negligible, see text). However, for A09, see panel (b), which is the fast-pulse, low-temperature case, larger separate connected structures are present, and these are color-coded.

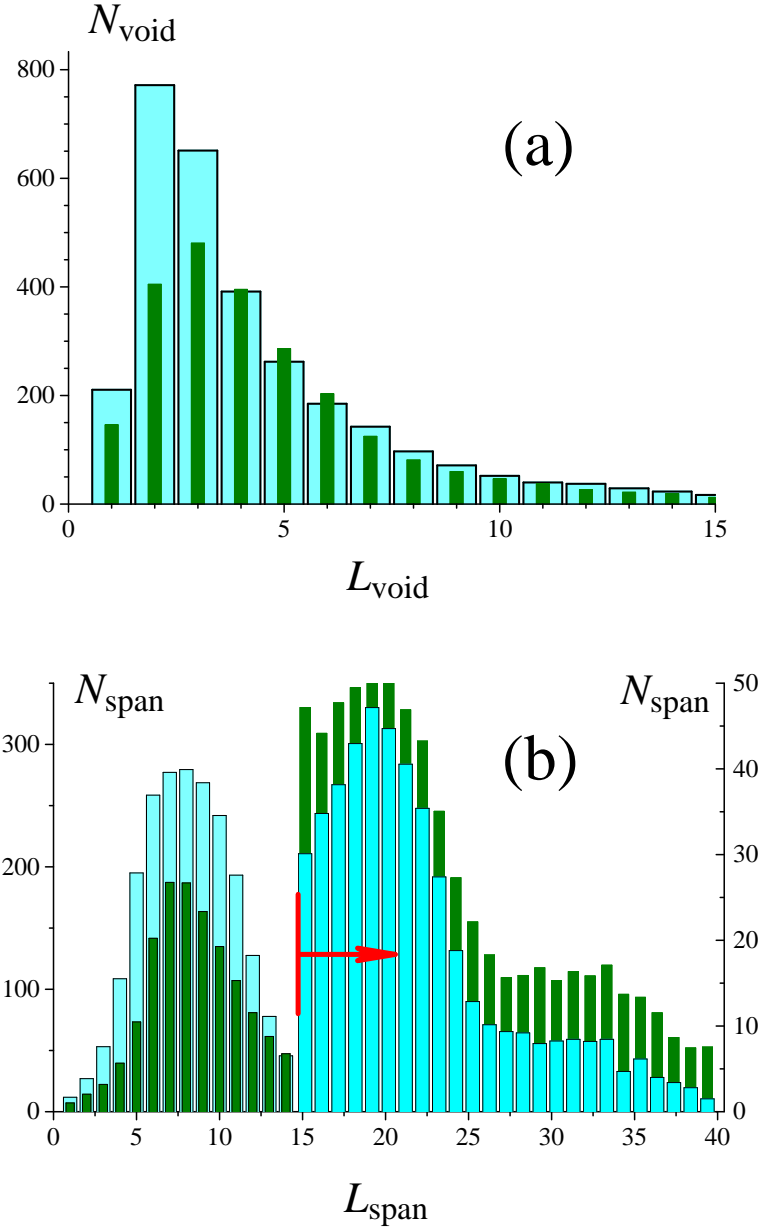

**Figure 8.** Comparison of the statistics of a measure of the quality of sintering for protocols A07 and A09. (a) Distribution of the lengths, $L_{void}$ of the segments of vacant lattice sites along the $x$ and $y$ directions in the $z = 0$ plane of the original particle centers. This measure of the void spans was summed up over all the relevant lattice rows, and averaged over 20 different MC runs (with random initial configurations). The bar coding: cyan for A09, green for A07. (b) The same for the segments of occupied lattice sites. This quantity measures the continuous in-plane spans, $L_{span}$, of the sintered structure along the $x$ and $y$ directions. The bar coding is the same as in (a). The arrow highlights span sizes well above those in the initial configuration, for which a different vertical scale is used (on the right).

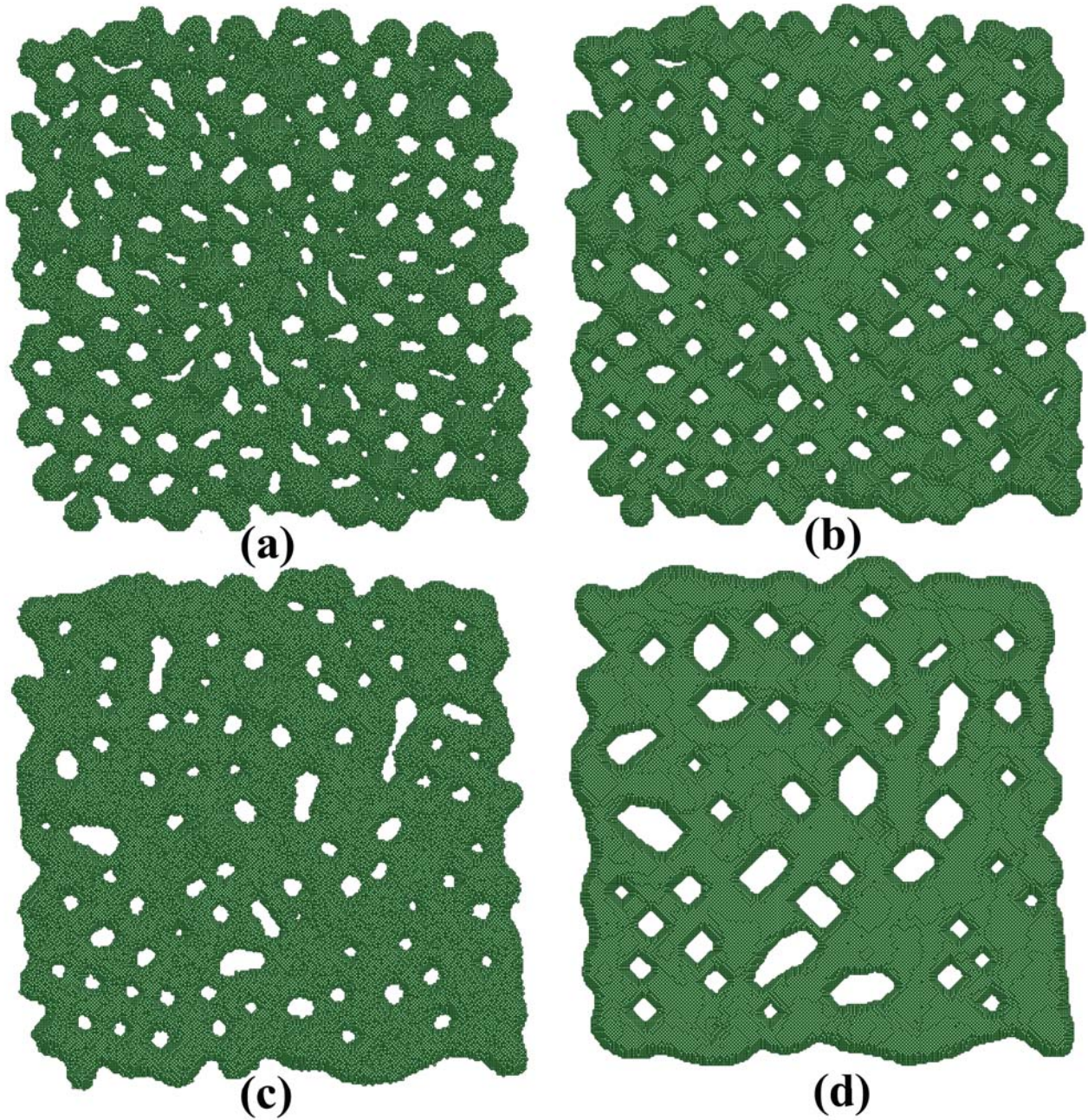

**Figure 9.** Time-dependence of the system starting with the Type II initial configuration shown in Fig. 2(b). Sintering with the temperature protocol A08 leads to the following configurations: (a) for $t = 8 \times 10^3$, at which time the maximum temperature for this protocol is reached, with $\alpha_{\min} = 0.8$; (b) for $t = 20 \times 10^3$, which is the ending time of the process (see Fig. 3). For the temperature protocol B08 we get the consecutive configurations (c) for $t = 50 \times 10^3$, which is the maximum-temperature time; (d) for $t = 150 \times 10^3$ time steps, which is the ending time.

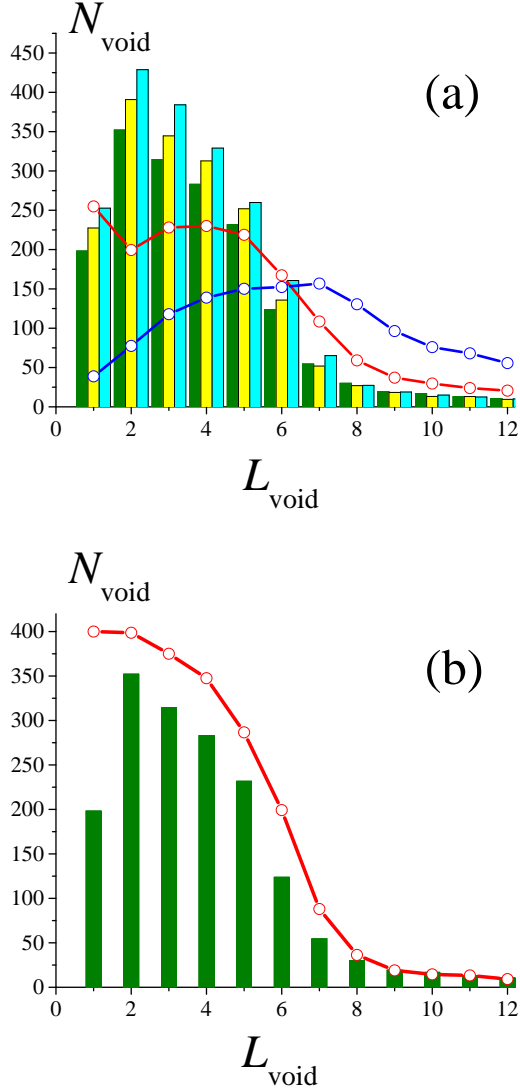

**Figure 10.** Similar to Fig. 8(a), but now for the Type II initial configurations, we consider the distribution of the lengths, $L_{\text{void}}$, of the segments of vacant lattice sites defined identically, and averaged over the same number of realizations. (a) The bars compare the final-time distribution for the fast-pulse temperature protocols with different maximum temperatures, color-coded: olive for A07, yellow for A08, cyan for A09. The open-circle symbols connected by color-coded segments to guide the eye, represent the $L_{\text{void}}$ statistics for the slow-pulse protocol B07 at two different process times: blue for the ending time of the process, and red for the time for which the maximum temperature is achieved. (b) The green bars are the same as in panel (a), and this distribution of the void sizes at the end of the fast protocol A07 is compared to the void-size counts at the time of the peak temperature for the same protocol, shown as the segment-connected circles.

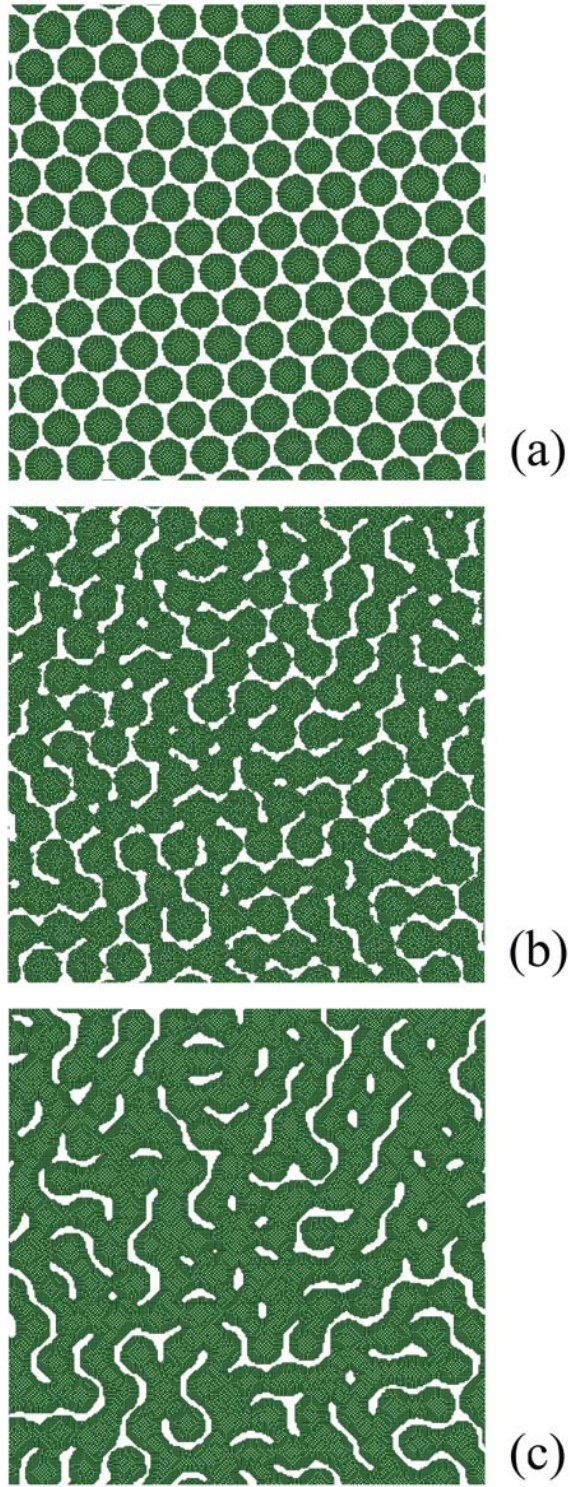

**Figure 11.** (a) Illustration of the initial configuration of uniform particles with minimal gaps approximately *a*. The dynamics of this system with temperature control protocol A08 is shown for (b) the time of the peak temperature, and (c) the ending time of the process.

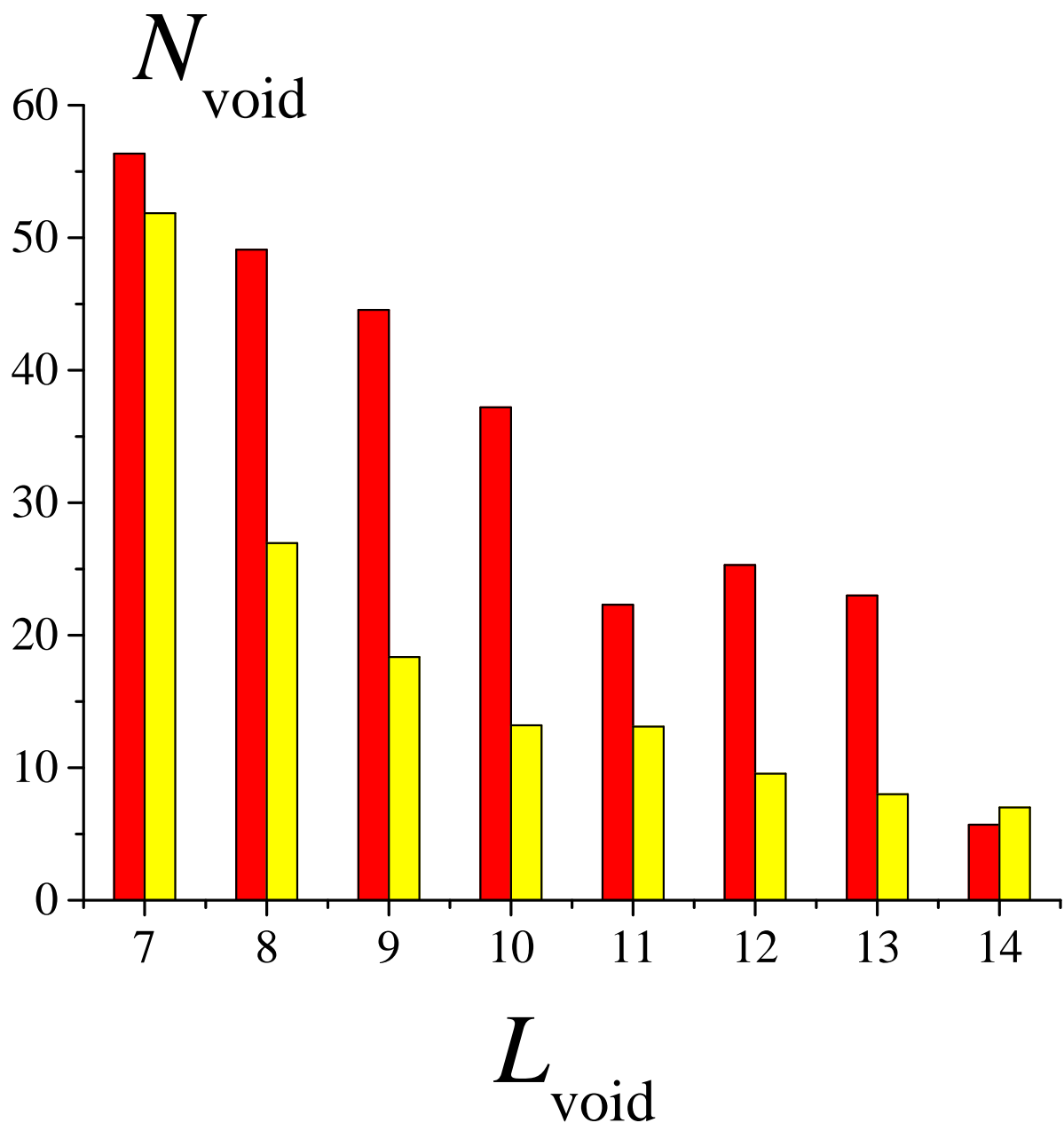

**Figure 12.** Comparison of the distribution of void sizes (except for the smallest ones) at the end of the sintering with the temperature-control protocol A08 for the uniform (red) and Gaussian (yellow) initial particle size distributions, the latter of Type II.

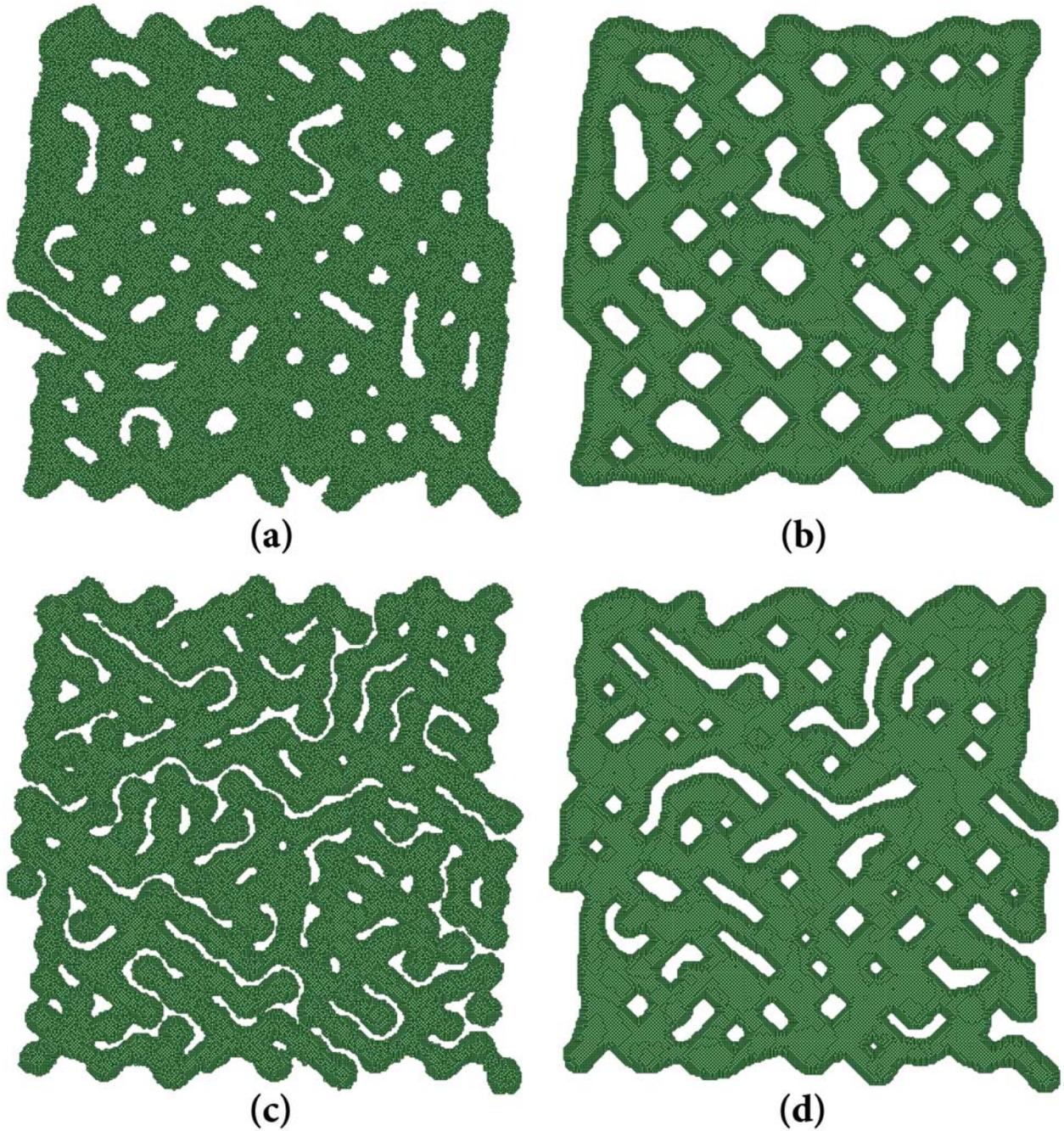

**Figure 13.** Typical realizations of the dynamics starting with a random initial state of uniform particles, different form that shown in Fig. 11(a), with a slow-pulse protocol B07, shown at (a) the peak temperature time, and (b) at the end of the process. Similar configurations obtained for the protocol B09 are shown in (c) peak-temperature time, and (d) ending time.

## Table of Contents Image (Graphical Abstract)

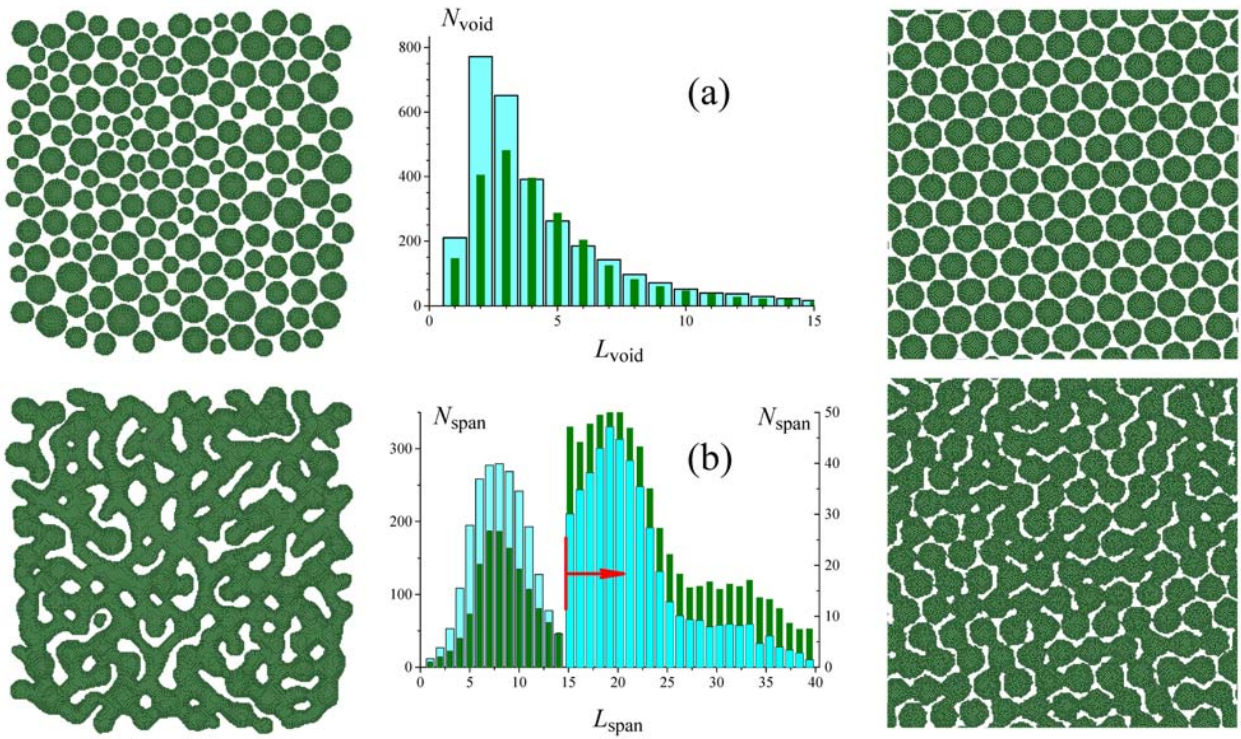

**Legend:** Sintering of layers of dispersed nanocrystals for different temperature control protocols and initial configurations.

# CrystEngComm

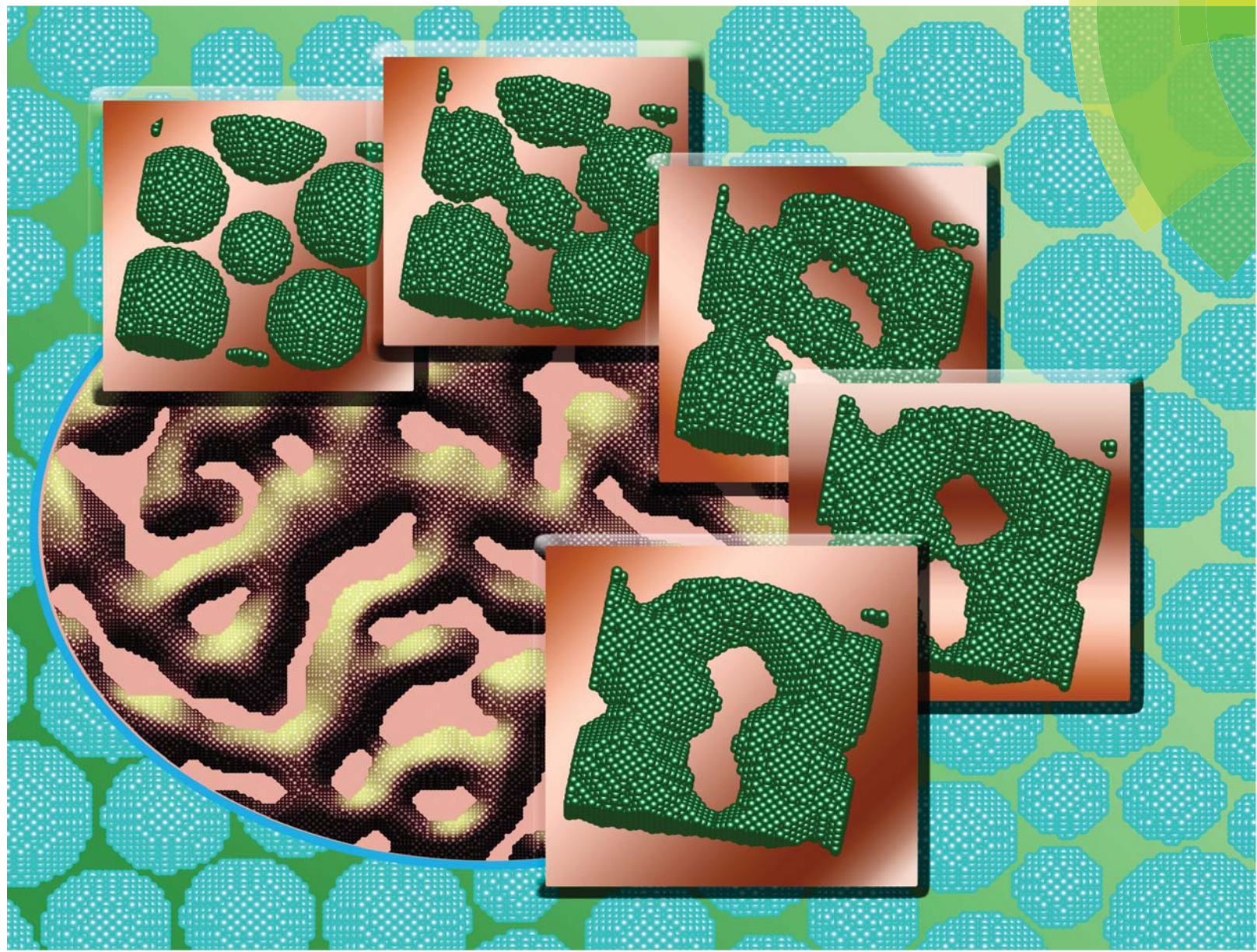

PAPER
Vladimir Privman et al.
Nonequilibrium kinetic modeling of sintering of a layer of dispersed nanocrystals